\documentclass[a4paper,12pt]{article}
\usepackage{amsmath}
\usepackage{color} \usepackage{amssymb} \usepackage{graphicx}

\begin{document}
  \newcommand{\bsg}{BR(b\rightarrow s \gamma)}
  \newcommand{\DM}{\Omega_{CDM}h^2}
  \newcommand{\gmu}{\delta a_{\mu}}
  \newcommand{\nn}{\nonumber}
  \newcommand{\stau}{\tilde{\tau}}
  \newcommand{\sel}{\tilde{e_R}}
  \newcommand{\smu}{\tilde{\mu_R}}
  \newcommand{\neut}{\tilde{\chi}^0_1}
  \newcommand{\nneut}{\tilde{\chi}^0_2}
  \newcommand{\nnneut}{\tilde{\chi}^0_3}
  \newcommand{\charg}{\tilde{\chi}^+_1}
  \newcommand{\ncharg}{\tilde{\chi}^+_2}
  \newcommand{\DeltaO}{\Delta^{\Omega}}

\begin{titlepage}


\begin{center}
{
\sffamily
\LARGE
Natural Implementation of Neutralino Dark Matter
}
\\[8mm]

S.~F.~King$^{a,b}$\footnote{E-mail: \texttt{sfk@hep.phys.soton.ac.uk}},
and J.~P.~Roberts$^b$\footnote{E-mail: \texttt{jpr@phys.soton.ac.uk}}
\\[5mm]
{\small\it
$^a$TH Division,Physics Department,\\
CERN, 1211,Geneva 23, Switzerland
}\\[5mm]
{\small\it
$^b$School of Physics and Astronomy,
University of Southampton,\\
Southampton, SO17 1BJ, U.K.
}\\[1mm]
\end{center}
\vspace*{0.75cm}

\begin{abstract}

\noindent
  The prediction of neutralino dark
  matter is generally regarded as one of the successes of the
  Minimal Supersymmetric Standard Model (MSSM). However the
  successful regions of parameter space allowed by WMAP and
  collider constraints are quite restricted. We discuss fine-tuning
  with respect to both dark matter and Electroweak Symmetry Breaking (EWSB) and
  explore regions of MSSM parameter space with non-universal
  gaugino and third family scalar masses in which neutralino dark
  matter may be implemented naturally.
  In particular allowing non-universal gauginos opens up the bulk region
  that allows Bino annihilation via t-channel slepton exchange,
  leading to ``supernatural dark matter''
  corresponding to no fine-tuning at all with respect to dark matter.
  By contrast we find that the recently proposed ``well
  tempered neutralino'' regions involve substantial fine-tuning
  of MSSM parameters in order to satisfy the dark
  matter constraints, although the fine tuning may be ameliorated
  if several annihilation channels act simultaneously. Although
  we have identified regions of ``supernatural dark matter'' in which
  there is no fine tuning to achieve successful dark matter, the
  usual MSSM fine tuning to achieve EWSB always remains.
\end{abstract}

\end{titlepage}
\newpage
\setcounter{footnote}{0}

  \section{Introduction}
  \label{Intro}

  One of the main arguments in favour of TeV scale Supersymmetry
  (SUSY) is that it provides a natural dark matter candidate
  \cite{hep-ph/0312378}.  In the minimal supersymmetric standard model
  (MSSM) with conserved R-parity the lightest supersymmetric particle
  (LSP) is absolutely stable.  If the LSP is the lightest neutralino
  it is also neutral, weakly interacting and has a mass of order the
  electroweak scale.  Although general arguments suggest that such a
  particle should provide a good dark matter candidate
  \cite{hep-ph/9506380}, the successful regions of parameter space
  allowed by WMAP and collider constraints are quite restricted
  \cite{hep-ph/0411216}-\cite{hep-ph/0402240}
  and it is far from clear if the neutralino can be considered a
  natural dark matter candidate.

  For example in the constrained MSSM (CMSSM)
  \cite{hep-ph/0411216} the majority of the parameter space
  results in a value of $\DM$ that exceeds the observed value by
  orders of magnitude. This is not to say that the CMSSM is ruled
  out by the WMAP measurement, just that it requires some very
  precise relations between parameters to fit the theory to the
  observed data. In the CMSSM the only region in which we can fit
  all experimental limits within $2\sigma$\footnote{This
  statement is true if one takes seriously the discrepancy
  between the reported value of the anomalous magnetic moment
  $g-2$ of the muon.  Throughout this paper we take the measured
  value of $\gmu$ that provides a $2.7\sigma$ deviation from the
  Standard Model \cite{hep-ex/0401008}.
  If this discrepancy were
  ignored, it would be possible to access the focus point and
  Higgs funnel regions of the CMSSM parameter space. However
  these turn out to be finely tuned regions as we shall discuss
  later.} is the $\stau$-coannihilation channel. In this region
  $m_{\stau}\approx m_{\neut}$ and the annihilation of SUSY
  particles in the early universe becomes extremely
  efficient. This effect is so large that the calculated value of
  $\DM$ drops by a factor of $100$ as $m_{\stau}$ approaches
  $m_{\neut}$. At one point on this steep gradient of decreasing
  $\DM$, the model gives the right amount of dark matter. However
  to be within this strip the low energy masses must be tuned to
  within a few percent. From a low energy point of view the MSSM
  contains no justification for the stau and neutralino masses to
  fall within this region and this has led to
  claims \cite{hep-ph/0601041} that such regions involve
  fine-tuning. Instead Arkani-Hamed, Delgado and Giudice argue
  that ``Well Tempered Neutralino'' regions in which the
  neutralino is not pure Bino but a Bino/Higgsino or Bino/Wino
  mix are more plausible.

  The present paper is concerned with the question of how to implement
  neutralino dark matter in a natural way within the general MSSM.
  Specifically we explore regions of MSSM parameter space with
  non-universal gaugino and third family scalar masses in which
  neutralino dark matter may be implemented naturally.  Dark matter
  with non-universal third family scalar masses has been considered in
  \cite{hep-ph/0403214} motivated by purely phenomenological
  consideration, or in \cite{hep-ph/0307389},\cite{hep-ph/0407165}
  motivated by specific GUT models. The collider phenomenology of such
  models has also been studied in \cite{hep-ph/9811300}. Dark Matter
  with non-universal gaugino masses has also been studied in
  \cite{hep-ph/0407218}-\cite{hep-ph/0102075} and particular high
  energy models with non-universal gauginos have been analysed in
  \cite{hep-ph/0603197}-\cite{hep-ph/0201001},\cite{hep-ph/0407165}.
  In both cases the question of naturalness of dark matter has not
  been addressed. Moreover no analysis of any kind has been performed
  which considers both types of non-universality together. Here we
  extend the above analyses considerably by studying the question of
  both dark matter and electroweak fine-tuning for both non-universal
  third family scalar masses and non-universal gauginos, including a
  first study of the effect of both types of non-universality
  together.

  Our main focus in this paper is on the question of whether and
  how naturalness may be improved by allowing such types of
  non-universality.  In order to examine the relative naturalness
  of different regions we employ a dark matter fine-tuning
  sensitivity parameter, which we use in conjunction with the
  similarly defined sensitivity parameter used for electroweak
  symmetry breaking (EWSB) \cite{Barbieri:1987fn},
  \cite{hep-ph/9810374}. Employing these quantitative measures of
  fine-tuning we find that $\stau$-coannihilation channel in the CMSSM may
  involve as little as 25\% tuning, due to renormalisation group
  (RG) running effects. This result is in agreement with
  \cite{hep-ph/0202110}.  In general moving beyond the framework of
  universal soft sfermion and gaugino masses in the CMSSM the
  structure of neutralino annihilation changes drastically. By
  allowing third family scalar masses to vary independently
  \cite{hep-ph/0403214}, \cite{hep-ph/0307389}, we can access
  regions in which the LSP has a significant Higgsino fraction as
  well as regions in which the dominant annihilation is through
  coannihilation with selectrons and smuons. With non-universal
  soft gaugino masses \cite{hep-ph/0407218} we can have a well
  tempered neutralino that is a Bino/Wino mix, although this is
  also very fine-tuned. However allowing non-universal gauginos
  also opens up a favourable region (excluded in the CMSSM)
  with light sleptons
  that allows Bino annihilation via t-channel slepton exchange.
  This yields a bulk region of parameter space leading to ``supernatural dark matter''
  where successful dark matter can be achieved with no
  fine-tuning at all, leaving only the usual fine-tuning required
  to achieve electroweak symmetry breaking.

  The paper is set out as follows.  Our methodology is stated in
  section \ref{methods}. In section \ref{bounds} we survey the
  physical bounds that constrain our parameter space, especially
  $\bsg$, $\gmu$ and $\DM$, we discuss radiative electroweak
  symmetry breaking, and we define the dark matter and EWSB
  sensitivity parameters that provide a measure of fine-tuning.
  In section \ref{CMSSM}, to set the scene, we consider the
  coannihilation region in the CMSSM, which provides us with a
  useful reference point against which the subsequent
  non-universal cases may be compared.  In section \ref{Scalar}
  we allow the third family soft sfermion mass squared to vary
  independently. In section \ref{Gauge} we consider neutralino
  dark matter with non-universal gaugino masses, but with a
  universal soft scalar mass.  In section
  \ref{ScalarGauge} we consider {\em both} the effects of
  including an independent third family sfermion mass squared
  {\em and} non-universal soft gaugino masses.
  Section \ref{Conc} concludes the paper.

  \section{Methodology}
  \label{methods}

  As the inputs for models such as the CMSSM are defined at
  $M_{GUT} \approx 2\times 10^{16}\ \text{GeV}$ we need to use the
  RGEs to produce a mass spectrum at the electroweak breaking
  scale. To do this we use {\tt SOFTSUSY
  v.1.9.1}\cite{hep-ph/0104145}. Once we have generated the low
  energy spectrum for a point we pass it to {\tt micrOMEGAs
  v.1.3.6}\cite{hep-ph/0112278} to calculate $\DM$, $\gmu$ and
  $\bsg$. We reject points that do not provide REWSB, that
  violate particle search limits from LEP2, points that produce a
  tachyon and any points that produce an LSP that is not the
  neutralino. In the remaining parameter space we plot the $1$
  and $2\sigma$ bounds of $\DM$, $\gmu$ and $\bsg$. Throughout we
  take the top mass to be $172.7\ \text{GeV}$.

  \section{Physical Bounds}
  \label{bounds}

  \subsection{$\bsg$}
  \label{bsgamma}

  The variation of $\bsg$ from the value predicted by the
  Standard Model is highly sensitive to SUSY contributions. To
  date no variation from the Standard Model has been
  detected. The present experimental measurement comes from BELLE
  \cite{hep-ex/0103042}, CLEO \cite{hep-ex/0108033} and ALEPH
  \cite{Barate:1998vz}. We follow the analysis of
  \cite{hep-ph/0403214} and take the value to be:
  \begin{equation}
    \bsg = (3.25 \pm 0.54) \times 10^{-4}
  \end{equation}
  The 1-loop SUSY processes involve loops with a charged Higgs
  and a top quark and loops with a chargino and a squark. Though
  the full 1-loop calculation is complicated, here we only
  mention that we expect contributions to be enhanced whenever
  the intermediate particles are light. There is also an
  enhancement for large $\tan\beta$.

  Within {\tt micrOMEGAs}, all 1-loop effects are included and
  some 2-loop contributions. There is a detailed discussion of
  their implementation in \cite{hep-ph/0112278}. We do not
  include theoretical errors in our analysis as they are hard to
  estimate at this stage.

  \subsection{Muon $g-2$}
  \label{gmu}

  Present measurements of the value of the anomalous magnetic
  moment of the muon $a_\mu$ deviate from the theoretical
  calculation of the SM value. However questions remain around
  the exact form of the standard model calculation, specifically
  whether we should use $\tau$ or $e^+e^-$ data in the analysis
  of the hadronic vacuum polarisation. At ICHEP
  '04\cite{hep-ph/0409360} there was general agreement that the
  $\tau$ data. Here we take the result
  obtained by using the $e^+e^-$ data and consider its
  implications for a SUSY theory. With the present experimental
  value from \cite{hep-ex/0401008} and the theoretical
  calculation of the SM value from \cite{hep-ph/0410081} there is
  a discrepancy:
  \begin{equation}
    (a_\mu)_{exp}-(a_\mu)_{SM}=\delta a_\mu = (2.52\pm
    0.92)\times 10^{-9}
  \end{equation}
  \noindent This amounts to a $2.7\sigma$ deviation from the
  standard model.\footnote{Note that the SND Collaboration recently reported
  a result that was out of line with the $e^+e^-$ results from
  other groups, being more
  consistent with the $\tau$ data. However very recently an error was
  reported in their analysis, and now the most recent result from SND
  is completely consistent with the $e^+e^-$ results from
  other groups \cite{SNP}. This effectively will serve to increase
  the discrepancy of the muon $g-2$ with the standard model
  beyond $2.7\sigma$, but since the new analysis has not
  yet been performed here we shall continue to conservatively assume the
  $2.7\sigma$ deviation.}

  The SUSY contributions to $a_\mu$ come from penguin diagrams of
  two types. One is mediated by a chargino and a muon sneutrino,
  the other is mediated by a neutralino and a smuon. For a
  detailed discussion of these contributions, see
  \cite{hep-ph/0103067}. For our purposes it is enough to note
  that there will be enhancements to the SUSY contribution
  whenever smuons, mu-sneutrinos, charginos and neutralinos
  become light.

  \subsection{$\DM$}
  \label{darkMatter}

  Evidence from the CMB and rotation curves of galaxies both
  point to a large amount of cold non-baryonic dark matter in the
  universe. The present measurements\cite{astro-ph/0302209} place
  the dark matter density at:
  \begin{equation}
    \DM = 0.1126 \pm 0.0081
  \end{equation}
  Due to R-parity, the lightest supersymmetric particle (LSP) is
  stable. If it is also electrically neutral, weakly interacting
  and massive it is a prime candidate for dark matter. In the
  majority of cases, just such a candidate exists in the
  neutralino. This is a mixture of the superpartners to U(1) and
  SU(2) gauge bosons of the standard model and the superpartners
  to the neutral Higgs bosons:
  \begin{equation}
    \nn \neut=N_{11}\tilde{B} + N_{12}\tilde{W} +
    N_{13}\tilde{H}_1^0 + N_{14}\tilde{H}_2^0
  \end{equation}
  \noindent where $N_{1,j}$ are the relevant components of the
  matrix that diagonalises the low energy neutralino mass matrix.

  This leads to three limiting cases:

    \begin{center}
      \begin{tabular}{ll}
    $\left|N_{11}\right|^2\approx 1$ & Bino dark
    matter\\ $\left|N_{12}\right|^2 \approx 1$ & Wino dark
    matter\\ $\left|N_{13}\right|^2 +
    \left|N_{14}\right|^2 \approx 1$  & Higgsino dark matter\\
      \end{tabular}
    \end{center}

  In many cases of pure Bino dark matter, t-channel slepton
  exchange is suppressed due to the sleptons being too heavy,
  resulting in a value of $\DM$ above the measured value. On the
  other hand pure Higgsino or Wino dark matter result in $\DM$
  being too small. This leads to the plausible suggestion of the
  ``well-tempered'' neutralino \cite{hep-ph/0601041} consisting
  of a roughly equal Bino/Wino or Bino/Higgsino mixed LSP. For
  example the well tempered Bino/Higgsino mixed LSP is achievable
  within the CMSSM in the so called Focus Point region where
  $m_0$ is large and $\mu$ is small \cite{hep-ph/0004043}. On the
  other hand the well tempered Bino/Wino mixed LSP is not
  achievable within the CMSSM, and requires non-universal gaugino
  masses.

  Alternatively pure Bino dark matter is still viable providing
  annihilation channels are enhanced for some reason. For example in
  the CMSSM the so called Coannihilation region is viable, where the
  Bino LSP is close in mass to the stau slepton leading to similar
  abundances of Binos and staus in the early universe, allowing efficient
  coannihilation into $Z$s and taus, for example, via t-channel neutralino
  exchange. Another example in the CMSSM is the so called Funnel region
  where $\tan \beta$ is large and the mass
  of the Bino LSP is equal to half the mass of the CP-odd pseudoscalar allowing
  efficient resonance annihilation.

  Both ``well tempered'' case and the enhanced Bino annihilation
  types of region are the exception rather than the rule. In
  the first case we need to tune the neutralino composition. In
  the second we need to tune the rest of the sparticle mass
  spectrum. The study of such tuning and how it may be overcome
  in certain regions of non-universal MSSM parameter space
  is the main subject of this paper.

\subsection{Radiative Electroweak Symmetry Breaking}

  One fundamental requirement of any SUSY theory is that it include
  radiative electroweak symmetry breaking (REWSB). For a detailed review of this
  process, see \cite{hep-ph/0312378}. Here we merely summarise the
  consequences of imposing REWSB. Firstly, by requiring REWSB, we swap
  the soft SUSY breaking parameters $\mu$ and $b$ for $\tan\beta$ and
  $\text{sign}(\mu)$. The size of $\mu^2$ is determined by:

  \begin{equation}
    \mu^2(t)=\frac{m^2_{H_d}(t)-m^2_{H_u}(t)\tan^2\beta}{\tan^2\beta-1}-\frac{1}{2}m_Z^2(t).
    \label{musq1}
  \end{equation}

  \noindent where $t=\log{Q}$ and $Q$ is the energy scale at which we
  want to determine $\mu$. For REWSB to exist in a given model, we must
  have a positive value of $\mu^2$ at the low energy scale $Q\sim m_Z$.

  $\tan\beta$ and $m_Z$ are easily determined at the low energy scale.
  $\tan\beta$ is considered as a free input for all cases considered and
  so just takes the value chosen for the model point under
  consideration. $m_Z$ is just the experimentally measured running mass
  of the $Z^0$. All the dependence of $\mu^2$ on the soft parameters is
  tied up in the $m_{H_u,H_d}^2$ terms. These masses must be evaluated
  at the low energy scale and so we must consider the RGE evolution from
  the soft scale to understand how varying the soft masses will affect
  $\mu^2$.

  It is worth noting
  that the $m_{H_d}^2$ term in Eq.~\ref{musq1} is suppressed relative to
  the $m_{H_u}^2$ term by a factor of $1/\tan^2\beta$. Therefore, for
  all but very small values of $\tan\beta$, the size of $\mu^2$ will be
  dominated by $m_{H_u}^2$. For large $\tan\beta$ (say larger than 2)
  Eq.~\ref{musq1} simplifies to:

  \begin{equation}
    \mu^2(t)\approx -m_{H_u}^2(t)-\frac{m_Z^2(t)}{2}
    \label{musimp}
  \end{equation}

  To obtain a positive value of $\mu^2$ in this limit we clearly need a
  negative value of $m_{H_u}^2$ at the low energy scale. As we have
  positive values for $m_{H_u,d}^2$ at $M_{GUT}$ in the models we
  consider, we must force $m_{H_u}^2$ negative through running
  effects.

  It is possible to calculate $\mu^2$ explicitly as a function of the
  high energy soft
  parameters \cite{hep-ph/9810374} (here we take $\tan\beta=10$ as we consider this
  region extensively throughout this work):

  \begin{eqnarray}
    \nn \frac{m_Z^2}{2}=&&-0.94\mu^2+0.010m_{H_1}^2-0.19M_2^2
    -0.0017M_1^2-0.63m_{H_2}^2+0.38m_{Q_3}^2\\
    \nn &&+0.38m_{U_3}^2
    +0.093A_t^2-0.011A_tM_1-0.070A_tM_2-0.30A_tM_3\\
    &&+2.51M_3^2+0.0059M_1M_2+0.028M_1M_3+0.195M_2M_3
    \label{musq}
  \end{eqnarray}

    This formula gives the condition amongst the
    high energy soft masses to achieve correct REWSB.
    It clearly shows that fine tuning is necessary if soft masses greatly
    exceed $m_Z$ (as they must). Correct REWSB clearly requires that the
    negative terms must not be too large compared to the positive terms, for
    example $m_{H_2}^2$ and $M_2^2$ must not be to large compared to
    $m_{Q_3}^2$ and $M_3^2$. Along the edge of regions where we fail to achieve REWSB, we
  will have regions with small $\mu$. These are favourable for two
  reasons. Firstly, such regions will facilitate neutralino annihilation
  via t-channel chargino exchange. Secondly, small $\mu$ means small
  fine-tuning of the $\mu$ parameter.

  \subsection{Fine-tuning sensitivity parameters}
  \label{finetuning}
  
  The naturalness of radiative electroweak symmetry breaking from
  supersymmetry has been extensively studied
  \cite{Barbieri:1987fn}-\cite{Allanach:2006jc}. In such studies, the
  following measure
is commonly used to quantify the degree of fine-tuning
  required for a given model point:

  \begin{equation}
    \Delta_a^{\text{EW}}=\frac{\partial \ln\left(m_Z^2\right)}{\partial
      \ln\left(a \right)}
    \label{EWmeas}
  \end{equation}

  \noindent where $a$ includes all the soft parameters
  $m_{\text{soft}}$ at the GUT scale together with $\tan \beta$. In
  this paper we wish to study the naturalness of dark matter. To this
  end we use an analagous measure to quantify the degree of fine-tuning
  required to reproduce the observed dark matter density:

  \begin{equation}
    \Delta_a^{\Omega}=\frac{\partial
    \ln\left(\DM\right)}{\partial \ln\left(a \right)}
    \label{meas}
  \end{equation}
  
  Though the measures are analagous, the calculations of radiative
  electroweak symmetry breaking and the calculation of the relic
  density of dark matter have important differences. In the case of
  electroweak symmetry breaking, the soft masses that define SUSY
  breaking provide a complete set of inputs. This is not the case for
  dark matter. The calculation of the present day relic density
  necessarily involves some assumptions about the cosmology of the
  early universe. These assumptions are:

  \begin{itemize}
  \item At some point in the universe's history (after inflation) there
    was a radiation dominated period in which $T \gg m_{\chi}$, where
    $m_{\chi}$ is the mass of the LSP.
  \item There are no exotic non-thermal production methods for dark matter.
  \end{itemize}

  If the first assumption holds then there was a period in the history
  of the universe in which standard model matter and supersymmetric
  matter (particularly the LSP) were in equilibrium. As the universe
  expands and cools a relic of stable SUSY particles is left
  behind. No further assumptions are required in the calculation
  of this relic density.
  
  If the second assumption holds then this LSP relic density is the
  only dark matter present in the universe. In this paper we work
  within such a cosmological framework. Within such a framework, the
  soft masses that parameterise the structure of SUSY breaking also
  uniquely define the relic density of dark matter. As a result, we
  can use Eq.\ref{meas} to quantify the naturalness of dark matter. In
  more exotic cosmologies, the set of input parameters $a$ would need
  to be expanded to include variations in the details of early
  universe cosmology but such cosmologies are beyond the scope of this
  paper.
  
  A similar measure to Eq.\ref{meas} was previously defined and used
  to study the CMSSM in \cite{Ellis:2001zk},\cite{hep-ph/0202110}. One
  motivation of these studies was to study the sensitivity that the
  LHC would need to achieve to give a corresponding accuracy in the
  predicted dark matter density.  This is an interesting question and
  our results can also be applied in this context, though we do not
  focus on that here.  In contrast to
  \cite{hep-ph/0202110}, we define the fine tuning of a model point to
  be equal to the largest individual tuning\footnote{There is a divide in the literature
    as to whether $\Delta=\text{max}(\Delta_a)$ or
    $\Delta=\sqrt{\sum_a \Delta_a^2}$. Which definition one takes to
    be more accurate is a matter of taste. Throughout the paper we
    take $\Delta=\text{max}(\Delta_a)$}: $\Delta^{\Omega}=
  \text{max}(\Delta_a^{\Omega})$.
  
  Note that the above sensitivity parameters in the electroweak sector
  have been criticized in
  \cite{Anderson:1995cp}-\cite{Anderson:1994dz} where an alternative
  measure of fine-tuning was proposed that arguably gives a more
  reliable estimate of fine-tuning. Essentially the authors of
  \cite{Anderson:1995cp}-\cite{Anderson:1994dz} point out that such
  sensitivity parameters do not provide a reliable measure of
  fine-tuning unless they are normalized to some average value of
  sensitivity for that particular parameter. Thus they advocate using
  instead the fine-tuning parameter $\gamma_a$ defined by,
  \begin{equation}
    \gamma_a = \frac{\Delta_a}{\overline{\Delta}_a},
    \label{gamma}
  \end{equation}
  where the average value ${\overline{\Delta}_a}$ is defined as
  discussed in \cite{Anderson:1995cp}-\cite{Anderson:1994dz}.  This
  definition in Eq.\ref{gamma} gives a sense to the numerical value of
  the sensitivity parameter, for example it may be the case that all
  values of the sensitivity parameter are large over the entire
  theoretically allowed range of input parameters. Fine tuning then
  corresponds to some unusually high levels of sensitivity above the
  typical values, where the typical values are themselves high.
  
  In the present paper it is sufficient to consider the unnormalized
  sensitivity parameters ${\Delta_a}$ for dark matter. For dark matter
  we shall find regions of parameter space where the sensitivities are
  very low, indeed sometimes less than unity. Conversely we shall find
  other regions of parameter space where the sensitivity parameters
  are very high, sometimes as high as of order $10^3$.  The existence
  of regions of parameter space with low sensitivities shows that the
  issue of the normalization of the sensitivity parameters is not so
  crucial in the case of dark matter.  Therefore we shall only
  calculate the absolute values of the sensitivity parameters which is
  sufficient to compare the relative sensitivies of different regions
  of parameter space.

  \section{CMSSM}
  \label{CMSSM}
  
  The simplest model of SUSY breaking is the CMSSM\footnote{See
    \cite{hep-ph/0411216} for a comprehensive discussion of the CMSSM
    parameter space. Here we merely present the most favourable region
    for the purpose of comparison with our later results.}. In such a
  model all soft mass matrices are diagonal at the high scale, taken
  here to be $M_{GUT}$, and have a universal mass squared, $m_0^2$,
  between generations.  The gauginos also have a universal soft mass
  $m_{1/2}$. After imposing REWSB we have 4 free parameters and a
  sign:

  \begin{center}
    $m_0$, $m_{1/2}$, $\tan\beta$, $A_0$ and sign$(\mu)$
  \end{center}
  
  Studies of the CMSSM find three regions that satisfy dark matter
  bounds. At high $m_0$, $\mu$ can become very small which in turn
  results in a Higgsino/Bino neutralino - the Focus Point region. At
  large $\tan\beta$ ($\approx 50$) and moderately large $m_{1/2}$ the
  pseudoscalar Higgs boson can become quite light giving rise to
  region in which $2m_{\neut}\approx m_{A}$. Here neutralino
  annihilation proceeds through s-channel exchange of a pseudoscalar
  Higgs boson, known as the Higgs Funnel. The third region that
  reproduces the observed value of $\DM$ is where we have
  coannihilation of the neutralino with the right handed stau. This
  region appears at low values of $m_0$ and $m_{1/2}$ where $m_{\stau}
  \approx m_{\neut}$. This region has a light SUSY spectrum and
  satisfies $\gmu$ and $\bsg$ at $1\sigma$.

  \begin{figure}
    \begin{center}
      \scalebox{.6}{\includegraphics{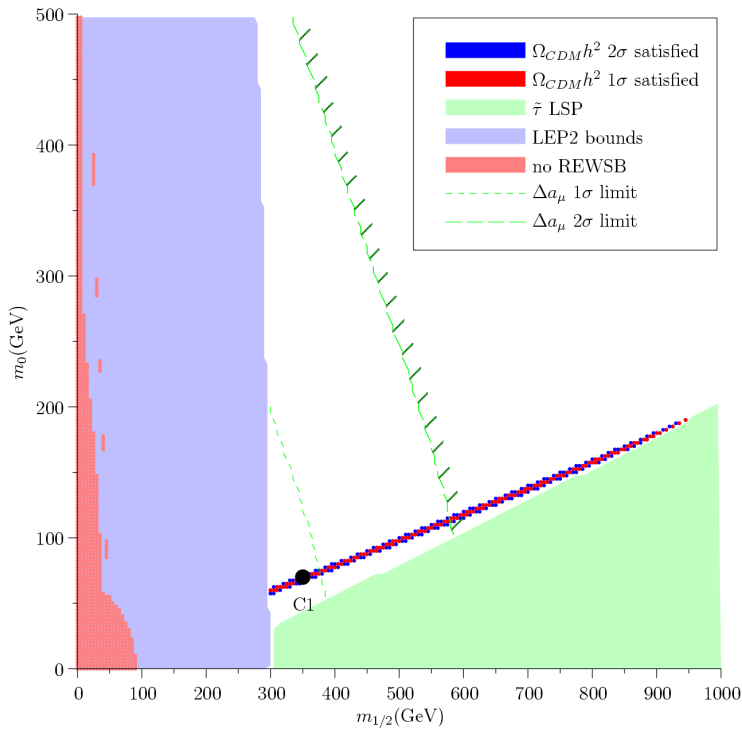}}
    \end{center}
    \caption{The $(m_{1/2},m_0)$
      plane for the CMSSM with $A_0=0$,
      $\tan\beta=10$.\label{fCMSSM}}
  \end{figure}
  
  In Fig.1 we study the CMSSM in the $(m_{1/2},m_0)$ plane with
  $A_0=0$, $\tan\beta=10$ and sign$(\mu)$ positive.  Low $m_{1/2}$
  values are excluded by LEP2 bounds on the lightest Higgs mass. Low
  $m_0$ values are excluded as they result in the stau becoming the
  LSP. In the remaining parameter space we plot the 1 and 2$\sigma$
  bounds on $\gmu$ and $\DM$. $\bsg$ is satisfied at $1\sigma$ through
  the entire parameter space and so does not appear as a bound. We
  have considered relatively low values of $m_0$ and $m_{1/2}$ as it
  is only in these regions that the CMSSM can satisfy $\gmu$.

  The region in which the CMSSM reproduces the observed value of
  $\DM$ is shown as a thin strip containing the point C1 running
  close the stau LSP light shaded region (the coannihilation
  strip is coloured red and blue for the $1\sigma$ and $2\sigma$
  regions respectively).  The coannihilation strip is the only
  viable dark matter region for the range of parameters in
  Fig.~\ref{fCMSSM}.  The thinness of the stau coannihilation
  strip indicates that some degree of tuning is required.  Before
  we go on to study how natural the region is, first we need to
  consider the neutralino annihilation channels that come into
  play in the calculation of the relic density.

  Through all the displayed parameter space the lightest
  neutralino is the Bino. Bino dark matter can always annihilate
  through the process $\neut\neut\rightarrow f\overline{f}$ via t
  and u-channel sfermion exchange where $f,\overline{f}$ are
  standard model fermions. This process becomes less efficient as
  the mass of the intermediate sfermions increase. However as low
  $m_0$ and $m_{1/2}$ are excluded, the sfermions of the CMSSM
  are never light enough for $\neut\neut\rightarrow
  f\overline{f}$ to account for the observed relic
  density. $\DM=\mathcal{O}(1)$ across the majority of the
  parameter space.


  To avoid overclosing the universe with an overabundance of dark
  matter we need another annihilation channel. This happens when
  $m_{\stau_R}\approx m_{\neut}$. In this case there is a
  significant number density of staus at the time of freeze
  out. This has two effects. Firstly, we must include processes
  of the form $\stau \stau \rightarrow \text{SM particles}$ and
  $\stau \neut \rightarrow \text{SM particles}$ in our
  calculation of the relic density of neutralinos. Secondly, it
  means that interactions will continue to occur long after
  neutralinos on their own would have frozen out\footnote{For a
  detailed study of coannihilation in general and the CMSSM
  specifically see \cite{Griest:1990kh} and \cite{hep-ph/9905481}
  respectively.}. The combination of effects results in a
  substantial decrease in the relic density. Unfortunately this
  generally leads to $\DM=\mathcal{O}(0.01)$. Nevertheless there
  is a band in which there is just enough coannihilation. This is
  just the coannihilation strip in Fig.~\ref{fCMSSM} that lies just
  above the region in which the $\stau$ is the LSP.

  As we move from one side of the strip to the other $\DM$ varies
  by roughly 2 orders of magnitude. Coannihilation will only
  provide the right amount of dark matter today if
  $m_{\stau}(m_Z)$ and $m_{\neut}(m_Z)$ are correlated to a
  precision of a few percent. In \cite{hep-ph/0601041}
  Arkani-Hamed, Delgado and Giudice use this critical
  sensitivity to the masses to claim that coannihilation requires
  fine-tuning. However we can only talk about fine-tuning in the
  context of a high energy model whereas their consideration of
  coannihilation is to consider the sensitivity on the low energy
  masses. To quantify the degree of fine-tuning in the
  coannihilation region, we take the point C1 with
  $m_0=70\ \text{GeV}$, $m_{1/2}=350\ \text{GeV}$, $\tan\beta=10$,
  $A_0=0$ and allow each input parameter to vary
  individually. Using the measure defined in Eq.~\ref{meas} we can
  quantify the sensitivity of the coannihilation strip to
  variations in each soft parameter at $m_{GUT}$, and this is done in
  Table~\ref{tCMSSM}.

  \begin{table}[ht]
    \begin{center}
      \begin{tabular}{|l|l|}
    \hline
    Parameter & Value \\
    \hline
    $\Delta_{m_0}^{\Omega}$ &       $3.5$\\
     $\Delta_{m_{1/2}}^{\Omega}$ &   $3.4$\\
    $\Delta_{\tan\beta}^{\Omega}$ & $1.4$\\
    $\Delta_{A_0}^{\Omega}$  &       $0$\\
    \hline
    $\Delta^{\Omega}$ & $3.5$\\
    \hline
    $\Delta^{\text{EW}}$ & $160$ \\
    \hline
      \end{tabular}
    \end{center}
    \caption{Fine tuning sensitivity parameters in the CMSSM with $A_0=0$,
    $\tan\beta=10$ at point C1 in Fig.~\ref{fCMSSM} with
    $m_0=70\ \text{GeV}$, $m_{1/2}=350\ \text{GeV}$.}
    \label{tCMSSM}
  \end{table}

  As coannihilation depends on the mass difference between the
  stau and the neutralino, we would expect the calculation of
  $\DM$ to be sensitive to $m_0$ and $m_{1/2}$. However the
  sensitivity of $\DM$ to these parameters is only around $25\%$,
  considerably less than would be expected from considerations of
  the low energy masses alone.  This is because along the
  coannihilation strip $m_0<m_{1/2}$ and so the masses of both
  the sleptons and the neutralino are primarily dependent on the
  same parameter $m_{1/2}$, in the case of the stau through RG
  running effects.

  Though here we only study coannihilation, there are other
  regions of the CMSSM parameter space in which we can reproduce
  the observed relic density, namely the Focus Point and Higgs
  Funnel regions. These appear at large $m_0$ and $\tan\beta$
  respectively and do not allow agreement with $\gmu$. However it
  is worth briefly mentioning that under the fine-tuning measure
  $\DeltaO$, the Focus Point region with Bino/Higgsino dark
  matter has $\DeltaO\approx60$ and the Higgs Funnel region has
  $\DeltaO\approx30$. Thus in the CMSSM, coannihilation provides
  more natural dark matter than the well-tempered neutralino of
  the Focus Point region.

  However this region does still exhibit fine-tuning. We now go
  on to consider models beyond the CMSSM to look for natural dark
  matter. We take point C1 as a datum to which we will
  compare regions that satisfy $\DM$ once we introduce varying
  degrees of non-universality in our soft parameters.

  \section{Non-universal Third Family Scalar Masses}
  \label{Scalar}

  The first deviation we take from universality of the CMSSM is
  to allow the 3rd family sfermion mass squared to vary
  independently of the 1st and 2nd families. This results in
  a model with five parameters and a sign:

  \begin{center}
    $m_0$, $m_{0,3}$, $m_{1/2}$, $\tan\beta$, $A_0$ and sign$(\mu)$
  \end{center}

  These determine the soft masses of the squarks, sleptons, Higgs
  and gauginos at $M_{GUT}$ to be:

  \begin{eqnarray*}
    m^2_{\tilde{Q}},m^2_{\tilde{L}},m^2_{\tilde{u}},m^2_{\tilde{d}},m^2_{\tilde{e}}&=&
    \begin{pmatrix}
      m^2_0 & 0 & 0 \\ 0 & m^2_0 & 0 \\ 0 & 0 & m^2_{0,3}
    \end{pmatrix} \\
    m^2_{H_u} = m^2_{H_d} &=& m_{0,3}^2 \\ M_\alpha &=& m_{1/2}
  \end{eqnarray*}

  \noindent where $\alpha=1,2,3$ labels the three gauginos. We
  have set the Higgs soft masses to be equal to the third family
  soft mass $m_{0,3}$ since it seems reasonable that all soft
  masses involved in EWSB should be of the same order.  Also this
  is the case in certain string models of non-universal third
  family scalar masses \cite{hep-ph/0211242},\cite{hep-ph/0403255}.

  From a purely phenomenological point of view, we gain a lot by
  allowing ourselves this extra freedom as pointed out in
  \cite{hep-ph/0403214}.  Firstly, the size of $\mu$ is primarily
  sensitive to the third family squark masses and the Higgs
  masses as shown in Eq.~\ref{musq}. On the other hand $\gmu$ is
  sensitive to the first and second family slepton masses. If we
  allow the 3rd generation soft masses to vary independently of
  the 1st and 2nd, we can access regions with low $\mu$ in which
  the neutralino is a Bino/Higgsino mix and still agree with
  $\gmu$ at $1\sigma$.

  However we have to be careful with low values of $\mu$. Small
  $\mu$ results in light charginos which enhance the SUSY
  contribution to $\bsg$. However the charginos appear in loops
  with stops, and with a large value of $m_{0,3}$, the stops
  become heavy and help to suppress the contribution. Thus in
  this model, $\bsg$ will be problem at large $m_{0,3}$ but not
  as much as might be initially expected.

  \begin{figure}
    \begin{center}
      \scalebox{.6}{\includegraphics{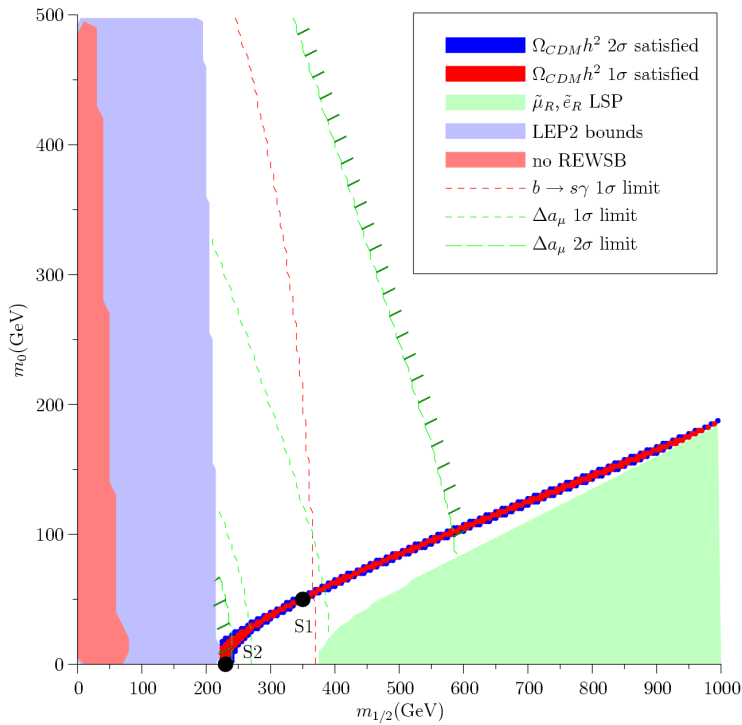}}
    \end{center}
    \caption{The $(m_{1/2},m_0)$ plane for non-universal sfermion
      masses with $m_{0,3}=1$TeV, $A_0=0$,
      $\tan\beta=10$. \label{fScalar1}}
  \end{figure}

  In Fig.~\ref{fScalar1} we take the same range for $m_0$ and
  $m_{1/2}$ as we did for Fig.~\ref{CMSSM} but we now set
  $m_{0,3}=1000\ \text{GeV}$. With a large value of $m_{0,3}$ the
  stau is no longer the lightest slepton, we now have a normal
  mass hierarchy (NMH) in the sfermions. The immediate result of
  this is that the coannihilation strip at low $m_0$ is now
  coannihilation with selectrons and smuons rather than
  staus. This also means that we can now access $m_0=0$ without
  ending up with a charged LSP. This happens at $m_{1/2}=230\ GeV$.
  The lower bound on $m_{1/2}$ from
  LEP2 constraints on the lightest Higgs also changes between
  Fig.~\ref{fCMSSM} and Fig.~\ref{fScalar1} from $300\ \text{GeV}$ to
  $m_{1/2}\approx 200\ \text{GeV}$.

  As we can access low values of both $m_{1/2}$ and $m_0$, we can
  access regions with light sleptons that in turn mediate the
  process $\neut\neut\rightarrow f\overline{f}$. As we move to
  low $m_0$ and $m_{1/2}$ this process becomes competitive with
  coannihilation. This shows up as a widening of the
  coannihilation region at low $m_0$ and $m_{1/2}$. However
  accessing this region comes at a cost. For low $m_0$ and
  $m_{1/2}<250\ \text{GeV}$ the smuons become so light that the
  SUSY contribution to $\gmu$ is too large. This large
  contribution could be reduced if we were to allow $M_1$ and
  $M_2$ to have different signs. Finally the dashed red line is
  also an early warning. For $m_{1/2}<350\ \text{GeV}$ the
  charginos are light enough that we violate the bounds
  on $\bsg$, though only at $1\sigma$.

  We consider two points on the successful dark matter strip:
  S1 at $m_0=50\ \text{GeV}$,
  $m_{1/2}=350\ \text{GeV}$; and S2 at $m_0=0\ \text{GeV}$,
  $m_{1/2}=230\ \text{GeV}$. The fine-tuning of these points with
  respect to our 5 parameters is shown in Table~\ref{tScalar1}.

  \begin{table}[ht]
    \begin{center}
      \begin{tabular}{|l|l|l|}
    \hline
    \multicolumn{1}{|c}{Parameter}&
    \multicolumn{1}{|c}{Value}&
    \multicolumn{1}{c|}{}\\
    \cline{2-3}
      &  S1 &         S2       \\
    \hline
    $\Delta_{m_0}^{\Omega}$ &       $2.4$ &      $0$  \\
    $\Delta_{m_{0,3}}^{\Omega}$ &   $0.15$ &     $0.30$  \\
    $\Delta_{m_{1/2}}^{\Omega}$ &   $4.2$ &      $1.8$  \\
    $\Delta_{\tan\beta}^{\Omega}$ & $0.061$ &    $0.033$ \\
    $\Delta_{A_0}^{\Omega}$ &       $0$ &        $0$ \\
    \hline
    $\Delta^{\Omega}$ & $4.2$ & $1.8$ \\
    \hline
    $\Delta^{\text{EW}}$ & $240$ & $200$ \\
    \hline
      \end{tabular}
      \caption{Fine tuning sensitivity parameters for different
        points in Fig.~\ref{fScalar1} with $m_{0,3}=1\ TeV$,
        $A_0=0$, $\tan \beta = 10$.  S1 has $m_0=50\ \text{GeV}$,
        $m_{1/2}=350\ \text{GeV}$. S2 has $m_0=0\ \text{GeV}$,
        $m_{1/2}=230\ \text{GeV}$ \label{tScalar1}}
    \end{center}
  \end{table}

  We choose point S1 to allow direct comparison with the
  CMSSM. Once again we find a tuning of around $25\%$ with
  respect to $m_{1/2}$ though the dependence on $m_{0,3}$ is
  minimal. This shouldn't be surprising as the neutralino
  annihilation channels in question depend primarily on the mass
  of the smuon, selectrons and the neutralino, none of which are
  sensitive to $m_{0,3}$. From this study, there is nothing to
  suggest that selectron and smuon coannihilation channels are
  any more natural that the stau coannihilation channel in the
  CMSSM.

  Point S2 is considerably more interesting. Here there is a
  dramatic decrease in the sensitivity to the soft
  parameters. The primary reason for this is that there are more
  channels at work than just coannihilation. At $m_0=0$, the
  selectron and smuon are light enough that t-channel slepton
  exchange in the process $\neut\neut\rightarrow e^+e^-\text{ or
  }\mu^+,\mu^-$ becomes competitive. Indeed at this point such
  processes account for $60\%$ of the annihilation of SUSY matter
  whereas coannihilation processes only account for $40\%$.

  This combination of annihilation channels is responsible for
  the drastic decrease in dark matter sensitivity to the soft
  parameters. By decreasing $m_{1/2}$ we decrease the mass of the
  neutralino and to lesser extent the selectron and the
  smuon. This increases the mass splitting between the states,
  suppressing coannihilation effects. However lower slepton
  masses enhance the cross-section for t-channel slepton
  exchange. This has the effect of smearing out the region of
  successful dark matter in the $m_{1/2}$ direction.

  If $m_0=0$, $\DeltaO_{m_0}=0$ automatically\footnote{Due to the
    definition of $\Delta^{\Omega}$, whenever $a=0$,
    $\Delta_a^{\Omega}=0$. However we have checked that this has not
    resulted in an artificially low $\Delta$ by taking a number of
    points of decreasing $m_0$. From this we conclude that as
    $m_0\rightarrow 0$, $\Delta_{m_0}^{\Omega}\rightarrow 0$
    smoothly.}. However we would also expect $\DeltaO_{m_0}$ to be
  small whenever $m_0$ is small. Whenever $m_0 \ll m_{1/2}$ the masses
  of both the sleptons and the neutralino are going to be primarily
  dependent upon $m_{1/2}$. Therefore the low energy phenomenology
  will be dominated by $m_{1/2}$ so neither t-channel slepton exchange
  or coannihilation should depend strongly on $m_0$ when $m_0$ is
  small.

  This particular combination of channels combine to smear out
  critical dependence on the soft parameters. Moreover, both
  channels will naturally occur in any model with light $m_0$ and
  reasonably light $m_{1/2}$. However light selectrons and smuons
  also enhance $\gmu$ and, as a result, point S2 slightly exceeds
  the $2\sigma$ bound on $\gmu$\footnote{As mentioned earlier
  this can be avoided if we allow a relative sign between $M_1$
  and $M_2$.}.

  Having access to $m_0=0$ also permits a solution to the SUSY
  flavour changing neutral current (FCNC) problem, at least for
  the first two generations, where it is most severe. Being zero
  at high energies, non-zero elements of the low energy upper
  block of the squark and slepton mass matrices proportional to
  the unit matrix are generated via gaugino running effects,
  giving a universal form involving the first two families,
  greatly suppressing the most severe FCNCs.  This mechanism is
  analogous to ``gaugino mediated SUSY breaking''
  \cite{hep-ph/0204108}, \cite{hep-ph/9911323} but applies here
  to only the first two families.  It was first studied in the
  framework of a particular brane set up in \cite{hep-ph/0012076} where
  it was referred to as ``brane mediated SUSY breaking''.

  \begin{figure}
    \begin{center}
      \scalebox{.6}{\includegraphics{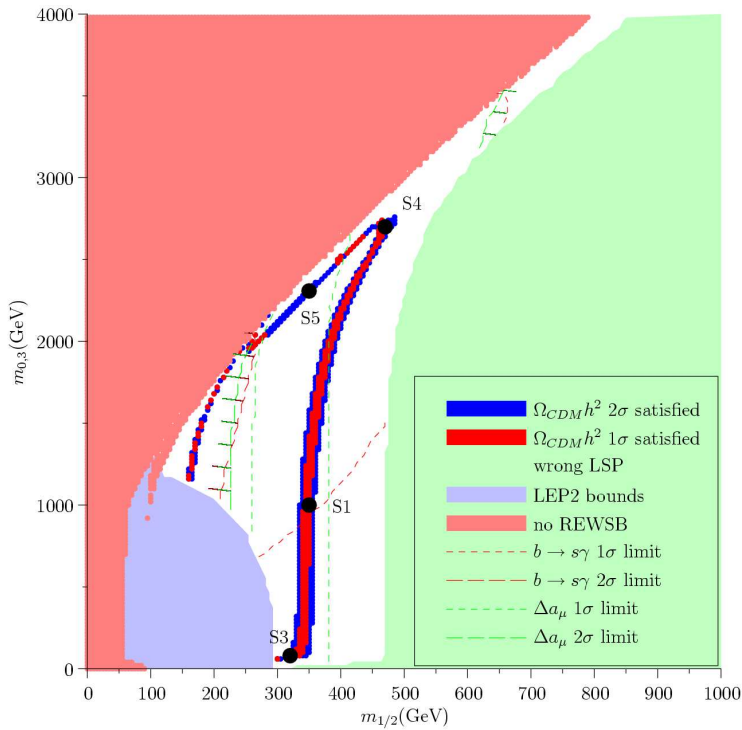}}
    \end{center}
    \caption{The $(m_{1/2},m_{0,3})$ plane for non-universal
      sfermion masses with $m_{0}=50$GeV, $A_0=0$,
      $\tan\beta=10$. \label{fScalar2}}
  \end{figure}

  Having considered the behaviour of a model with
  $m_{0,3}=1\text{TeV}$, we now go on to consider the case of
  general $m_{0,3}$ in Fig.~\ref{fScalar2}. We fix
  $m_0=50\ \text{GeV}$ and allow $m_{0,3}$ to vary from
  $1-4\text{TeV}$. For the majority of the
  plot, the light shaded (green) region represents the region ruled out
  due to a smuon and selectron LSP. The dark matter strip that
  runs parallel to it is due to coannihilation with smuons and
  selectrons, as exemplified by point S1 in Fig.~\ref{fScalar2}
  also studied previously.

  As we mentioned earlier, if we increase $m_{0,3}$ enough we
  lower $\mu^2$ to the point where it becomes negative and REWSB
  fails. As we approach the region in which REWSB fails, $|\mu|$
  decreases and therefore we steadily increase the Higgsino
  component of the neutralino. This results in an analogue of the
  Focus Point region in the CMSSM. As well as an LSP with a
  significant Higgsino component we also have light
  charginos. These two factors together enhance decays of the
  form $\neut\neut\rightarrow W^+W^-$. $\nneut$ and $\nnneut$
  also pick up substantial Higgsino components and thus become
  close in mass to $\neut$, allowing another form of
  coannihilation. All of these channels are most efficient when
  $\mu<M_1$ and the neutralino is pure Higgsino. In such cases
  the annihilation of SUSY matter in the early universe is so
  efficient that we would end up with no appreciable dark matter
  relic today. Therefore agreement with the observed relic
  density is achieved when the neutralino is a
  well tempered Bino/Higgsino mix, as in point S5 in
  Fig.~\ref{fScalar2}.

  The last band in Fig.~\ref{fScalar2} that agrees with $\DM$ is
  at low $m_{1/2}$ and $m_{0,3}\approx 1.2-2\text{TeV}$. In this
  region $2m_{\neut}\approx m_{h}$ and annihilation proceeds
  through resonant s-channel exchange of a light Higgs. However
  this region is ruled out at $2\sigma$ by $\bsg$ and $\gmu$ so
  we do not study it further here.

  \begin{table}[ht]
    \begin{center}
      \begin{tabular}{|l|l|l|l|}
    \hline
    \multicolumn{1}{|c}{Parameter}&
    \multicolumn{1}{|c}{Value}&
    \multicolumn{1}{c}{}&
    \multicolumn{1}{c|}{}\\
    \cline{2-4}
     & S3 &          S4 &        S5 \\
    \hline
    $\Delta_{m_0}^{\Omega}$ &       $1.5$ &      $0.67$ &     $0.12$ \\
    $\Delta_{m_{0,3}}^{\Omega}$ &   $0.41$ &     $2.4$ &      $30$ \\
    $\Delta_{m_{1/2}}^{\Omega}$ &   $2.4$ &      $2.8$ &      $18$ \\
    $\Delta_{\tan\beta}^{\Omega}$ & $0.23$ &     $1.0$ &     $12$ \\
    $\Delta_{A_0}^{\Omega}$ &       $0$ &    $0$ &    $0$ \\
    \hline
    $\Delta^{\Omega}$ & $2.4$ & $2.8$ &     $30$ \\
    \hline
    $\Delta^{\text{EW}}$ & $91$ & $1300$ &  $950$ \\
    \hline
      \end{tabular}
      \caption{Fine tuning sensitivity parameters
      at points in Fig.~\ref{fScalar2}. S3 has $m_{0,3}=80\ \text{GeV}$,
        $m_{1/2}=320\ \text{GeV}$. S4 has $m_{0,3}=2700\ \text{GeV}$,
        $m_{1/2}=470\ \text{GeV}$. S5 has $m_{0,3}=2308\ \text{GeV}$,
        $m_{1/2}=350\ \text{GeV}$ \label{tScalar2}}
    \end{center}
  \end{table}

  To study the naturalness of the dark matter regions, we take
  the points S3-S5 as shown in Fig.~\ref{fScalar2} and
  Table~\ref{tScalar2}.  Point S1 also shown in
  Fig.~\ref{fScalar2} was considered previously.  This allows a
  comparison of the Focus Point region with its well-tempered
  neutralino in point S5 to the smuon/selectron coannihilation
  region of point S1. We also take points S3 and S4 where these
  regions meet. Having found a hint of natural dark matter in
  point S2 in which we had more than one annihilation channel at
  work, S3 and S4 allow us to study the effect of regions that
  exhibit different combinations of annihilation channels.

  Point S3 lies on the point at which selectrons, smuons and
  staus all contribute to annihilation rates equally. We also
  have a $47\%$ contribution from t-channel slepton
  exchange. Because the stau coannihilation depends almost solely
  on $m_{0,3}$ and the selectron and smuon coannihilation depends
  on $m_0$, the sensitivity to either of these parameters is
  reduced. In addition, if we increase either $m_{0,3}$ or $m_0$
  we are increasing the mass of either the staus or the smuons
  and selectrons, which always leaves at least one light slepton
  to mediate t-channel slepton exchange. As a result S3 has a
  lower fine tuning than either primarily
  stau coannihilation (as with C1) or selectron and smuon
  coannihilation (as with S1).

  Point S4 also lies on an intersection in annihilation
  channels. This time it is the intersection of the Higgsino/Bino
  ``well-tempered'' strip and the selectron, smuon coannihilation
  strip. The quoted $\Delta_a^{\Omega}$ values should therefore
  be compared to S5 and S1 respectively. The selectron, smuon
  coannihilation strip is primarily sensitive to $m_0$ and
  $m_{1/2}$ through their effect on the sparticle masses. In
  contrast the Higgsino/Bino strip, represented by point S5, is
  highly sensitive to $m_{0,3}$ and $m_{1/2}$ through their
  effects on the low energy values of $\mu$ and $M_1$. Once again
  by combining channels we reduce our dependence on any one
  parameter. In this case we manage to achieve a low degree of
  fine-tuning while also having a well-tempered
  neutralino. However with $m_{0,3}=2700\ \text{GeV}$ we are in
  peril of reintroducing fine-tuning complaints in the Higgs
  sector.

  Finally, note that the least natural region we have considered
  is that of the Higgsino/Bino neutralino without any other
  annihilation channels. In \cite{hep-ph/0601041} the
  well-tempered neutralino is suggested as some of the most
  plausible options for a general SUSY theory. However here we
  have shown that, at least in certain models, such well-tempered
  regions are more fine-tuned with respect to soft parameters
  than coannihilation regions.

  In summary, allowing $m_{0,3}$ to vary independently of the first and second
  family masses, and in particular to become independently large,
  we find the following features:
  \begin{itemize}
  \item Access to $m_0=0$, and hence a solution to the SUSY FCNC problem,
   with lower $\Delta_a^{\Omega}$ than in the CMSSM at the
    expense of large SUSY contributions to $(g-2)_\mu$.
  \item Access to the well tempered Higgsino/Bino region which agrees with $\gmu$
    at $1\sigma$, but this well tempered point involves $3\%$ dark matter fine tuning.
  \item We can access certain regions such as S3,S4 in which we have a number of
    different annihilation channels at work and these lead to
    lower values of $\Delta_a^{\Omega}$ and more natural solutions to the dark
    matter problem.
  \item Large $m_{0,3}$ implies a high degree of fine tuning for REWSB, as expected,
  so a fully natural model is not possible in this case.
  \end{itemize}

  \section{Non-universal Gaugino Masses}
  \label{Gauge}

  As with high 3rd family masses, there are good reasons for
  allowing the soft gaugino masses to be non-universal. From a
  theoretical point of view, non-universal gaugino masses are
  interesting as they naturally occur in SUGRA models with
  non-minimal gauge kinetic terms, which arise from many string
  constructions (see e.g. \cite{hep-ph/0403255}, \cite{hep-ph/0211242}).
  They also naturally occur in gauge mediated SUSY breaking and
  anomaly mediated SUSY breaking, for example
  \cite{hep-ph/9607225}, \cite{hep-ph/9907319} respectively.

  From a phenomenological point of view, non-universal gauginos
  are extremely useful. By allowing $M_1$ and $M_2$ to take
  different values, we can directly change the Bino/Wino balance
  in the neutralinos allowing access to Wino and well tempered
  Bino/Wino LSP states. $M_3$ allows us to vary the squark sector
  independently of the sleptons through its effects on the
  running masses. $\mu$ also depends on $M_3$ as shown in
  Eq.~\ref{musq}.  Therefore by allowing our gaugino masses to be
  non-universal we gain control over $M_1$, $M_2$ and to a lesser
  extent $\mu$.

  \begin{figure}
    \begin{center}
      \scalebox{.6}{\includegraphics{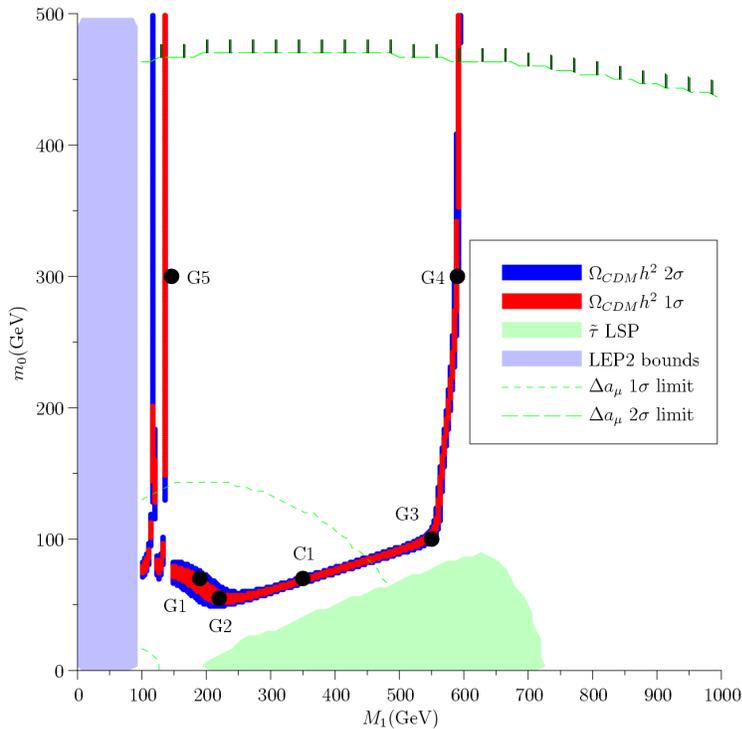}}
    \end{center}
    \caption{The $(M_1,m_0)$ plane for non-universal gauginos
      with $M_2=M_3=350$GeV, $A_0=0$,
      $\tan\beta=10$.\label{fGauge1}}
  \end{figure}

  To be precise we consider the parameters:
  \begin{center}
    $m_0$, $M_1$, $M_2$, $M_3$, $\tan\beta$, $A_0$ and sign$(\mu)$
  \end{center}

  In Fig.~\ref{fGauge1} we keep $M_2=M_3=350\ \text{GeV}$,
  $\tan\beta=10$ and $A_0=0$, but allow $m_0$ and $M_1$ to vary. As we
  have a unified mass for the sfermions, the light shaded (green) region
  at low $m_0$ is excluded by a stau LSP. The LEP2
  bound at $M_1\approx 100\ \text{GeV}$ is due to the neutralino becoming
  too light. With $M_3=350\ \text{GeV}$ the Higgs mass is
  above the LEP limit across the whole scan\footnote{In parallel to
  Fig.~\ref{fCMSSM}, we can take $M_3$ down to $300\ \text{GeV}$
  without violating the LEP2 bound on the lightest Higgs. This
  also results in the lowest value of $\mu$.}.

  The most striking feature of Fig.~\ref{fGauge1} is the multitude
  of different regions that satisfy $\DM$. There is the
  coannihilation strip for $300\ \text{GeV} < M_1 < 500\ \text{GeV}$
  containing the point C1 at $M_1=350\ GeV$ which coincides with
  the CMSSM point considered previously, with Bino dark matter
  and a stau coannihilation channel.  In addition there is the
  well tempered Bino/Wino LSP at $M_1\approx 600 \ GeV$
  containing the point G4. This value of $M_1$ corresponds to the
  high energy relation $M_1\approx 1.7 M_2$ corresponding to the
  low energy equality $M_1(m_Z)\approx M_2(m_Z)$.  The two
  vertical lines at $M_1=130\ \text{GeV}$ and $M_1=140\ \text{GeV}$
  correspond to the points at which $2m_{\neut}=m_Z,m_h$
  respectively and we have resonant s-channel annihilation.
  However the most interesting region is the bulk region
  containing the point G1 which arises from the very low values
  of $150<M_1<250$ GeV which are allowed now that gaugino
  universality is relaxed, which allows the sleptons to be light
  enough that $\neut\neut\rightarrow l\overline{l}$ can produce
  the observed relic density on its own. We study fine tuning in
  these regions by calculating $\Delta_a^{\Omega}$ for points
  C1,G4,G5,G1 in each region and for points G2,G3 where two
  annihilation processes contribute equally. The results are
  presented in Table~\ref{tGauge1}.
  \footnote{Though point C1 has the same values for
    the soft parameters as in Table~\ref{tCMSSM}, here we allow the
    gaugino masses to vary independently. Therefore, though it is
    labelled C1, it does not represent the CMSSM. This is the point at
    which a model with non-universal gauginos makes contact with the CMSSM.}

  \begin{table}[ht]
    \begin{center}
      \begin{tabular}{|l|l|l|l|l|l|l|}
    \hline
    \multicolumn{1}{|c}{Parameter}&
    \multicolumn{1}{|c}{Value}&
    \multicolumn{1}{c}{}&
    \multicolumn{1}{c}{}&
    \multicolumn{1}{c}{}&
    \multicolumn{1}{c}{}&
    \multicolumn{1}{c|}{}\\
    \cline{2-7}
     &                              G1 &      G2 &      G3 &     G4 &
    G5 &        C1   \\
    \hline
    $\Delta_{m_0}^{\Omega}$ &       $0.83$ &  $0.97$ &  $3.0$ &  $0.65$ &   $5.7$    &  $3.5$  \\
    $\Delta_{M_1}^{\Omega}$ &       $0.80$ &  $0.51$ &  $8.0$ &  $28$   &   $1100$    & $2.7$  \\
    $\Delta_{M_2}^{\Omega}$ &       $0.23$ &  $0.36$ &  $3.4$ &  $26$   &   $4.8$    &  $0.64$ \\
    $\Delta_{M_3}^{\Omega}$ &       $0.24$ &  $0.44$ &  $2.3$ &  $5.8$   &  $91$    &   $1.4$  \\
    $\Delta_{\tan\beta}^{\Omega}$ & $0.20$ &  $0.50$ &  $1.2$ &  $0.20$ &   $4.1$    &  $1.4$  \\
    $\Delta_{A_0}^{\Omega}$ &       $0$ &     $0$ &     $0$ &    $0$  & $0$  &  $0$\\
    \hline
    $\Delta^{\Omega}$ & $0.83$ & $0.97$ & $8.0$ &    $28$ &     $1100$ &    $3.5$ \\
    \hline
    $\Delta^{\text{EW}}$ & $110$ & $110$ & $111$ &   $111$ &    $110$ &     $160$ \\
    \hline
      \end{tabular}
      \caption{Fine tuning sensitivity parameters for
    points taken from Fig.~\ref{fGauge1} with $M_2=M_3=350$GeV, $A_0=0$,
      $\tan\beta=10$. G1
    has $m_0=70\ \text{GeV}$, $M_1=190\ \text{GeV}$. G2 has
    $m_0=55\ \text{GeV}$, $M_1=220\ \text{GeV}$. G3 has
    $m_0=100\ \text{GeV}$, $M_1=550\ \text{GeV}$ G4 has
    $m_0=300\ \text{GeV}$, $M_1=590\ \text{GeV}$ G5 has
    $m_0=300\ \text{GeV}$,
    $M_1=146.15\ \text{GeV}$.
    \label{tGauge1}}
    \end{center}
  \end{table}

  It is striking that two of the points in Table~\ref{tGauge1},
  namely G1 and G2, have a dark matter sensitivity parameter
  below unity, corresponding to ``supernatural dark matter''
  (i.e. no fine tuning required at all to achieve successful dark
  matter).  Point G1 lies in the middle of the bulk region in
  which annihilation proceeds through t-channel sfermion
  exchange. As this process depends directly upon the mass of the
  sleptons that are exchanged, it is unsurprising that the
  majority of the dependence is on $m_0$ and $M_1$. However it is
  clear that the dependence is significantly lower than the
  CMSSM. This is because the cross-section for
  $\neut\neut\rightarrow l\overline{l}$ varies slowly with
  $m_{\tilde{l}}$ in comparison to coannihilation processes that
  depend upon precise mass differences between particles. In the
  neighbouring point G2 both coannihilation and t-channel slepton
  exchange are competitive. Here the sensitivity to $m_0$
  increases slightly (due to the coannihilation channel) but the
  sensitivity to $M_1$ decreases significantly for the same
  reasons as point S2 in the the case of non-universal scalars.

  The advantage of combining annihilation channels is shown
  by comparing the point G3 and G4. G4 has been chosen to
  lie on the well-tempered Bino/Wino line. As with the Bino/Higgsino
  region, as we move away from a pure Bino LSP the annihilation
  and coannihilation cross-section rise dramatically, leading to
  a sharp drop in $\DM$. Here, as before, such ``well-tempered''
  regions exhibit $\Delta_a^{\Omega}$ values of $\approx30$, values well in excess
  of slepton coannihilation regions.
  G3 lies in the region in which we have both coannihilation with
  staus and a significant Wino component in the neutralino. This
  results in a region that is both ``well-tempered'' and exhibits
  lower values of $\Delta_a^{\Omega}$.

  Finally we briefly consider the naturalness of s-channel
  annihilation regions. As these are sharply peaked whenever
  $2m_{\neut}=m_{Z,h}$, we expect there to be a substantial
  dependence on any parameter that directly affect the neutralino
  mass. As point G5 shows, this dependence is extreme. If we were to
  find ourselves to be living in this part of parameter space we
  would have to look for some further theoretical
  justification. An $\Delta_a^{\Omega}$ value greater than 1000 cannot be
  considered natural.

  Once again, these studies of naturalness can only be considered
  in the case of a given model. In this model with non-universal
  gaugino masses we have showed that the ``well-tempered'' region
  is considerably more fine tuned than t-channel slepton exchange
  or even slepton coannihilation regions. However the tuning in
  the ``well-tempered'' region pales in significance when we
  consider the amount of tuning required to land us on the edge
  of the light Higgs annihilation channel.

  However the fate of the Wino/Bino dark matter strip could be
  considered to be better than that of the Bino/Higgsino
  neutralino. The reason for this is that the Bino/Higgsino
  region depends precisely on $M_1/\mu$ whereas the Bino/Wino
  region depends on $M_1/M_2$. In the latter case to guarantee
  that we land in the Bino/Wino well-tempered strip it is only
  necessary to have a model that requires $M_1(m_{GUT})\approx
  1.7 M_2(m_{GUT})$. In such a model the large dependence on $M_1$
  and $M_2$ in G4 would disappear and a Wino/Bino neutralino could
  provide a dark matter candidate. However there is no magic
  ratio at the high scale that will obviously reproduce the
  relevant Higgsino/Bino ratio at the low scale. To see this it
  is enough to consider the dependence of $\mu$ on the soft
  parameters. To guarantee a given ratio between $\mu$ and $M_1$
  it would be necessary to have a model that relates all the
  parameters in Eq.~\ref{musq} in just the right way.

  Leaving aside such questions of models of SUSY breaking, the
  most vivid result of allowing non-universal gaugino masses are:

  \begin{itemize}
  \item The bulk region involving Bino annihilation via
    t-channel slepton exchange can be accessed, since low $M_1$
    is possible leading to light sleptons. This allows
    supernatural dark matter exemplified by the points G1,G2.
  \item Low $m_0$ and $M_1$ also allow us to fit $\gmu$ at
    $1\sigma$.
  \item A well tempered Bino/Wino neutralino scenario is
    possible, for example point G4, but this requires $3\%$ fine
    tuning as in the case of the well tempered Bino/Higgsino.
  \item However combinations of annihilation channels on the edge
    of such regions such as point G3 lead to lower fine tuning.
  \item To achieve the observed value of $\DM$ through resonant
    annihilation such as at point G5 through a light Higgs
    requires extremely large fine tuning.
  \end{itemize}

  \section{Non-universal Gauginos and Third Family Scalar Masses}
  \label{ScalarGauge}

  Having considered non-universal gauginos and the situation in
  which we allow the third family sfermion and Higgs masses to be
  large, we now consider the effect of including both of these
  extensions to the CMSSM at once. This results in a model with 7
  free soft parameters and a sign:

  \begin{center}
    $m_0$, $m_{0,3}$, $M_1$, $M_2$, $M_3$, $\tan\beta$, $A_0$ and sign$(\mu)$
  \end{center}

  It is clear from the two previous sections that the mechanics
  of dark matter annihilation can be sensitive to many of these
  parameters. To get a true handle of the sensitivity of these
  regions to the soft parameters, we should allow all of the
  different parameters to vary at once. In addition, by allowing
  all parameters to vary simultaneously, we open up the
  possibility of accessing new regions in which we satisfy dark
  matter that have even lower $\Delta_a^{\Omega}$ values than we
  have found so far. However having large $m_{0,3}$ clearly means
  that REWSB will be fine tuned.  Nevertheless a new result of
  having both types of non-universality together is that we can
  access a maximally tempered Bino/Wino/Higgsino neutralino where
  the LSP consists of roughly equal amounts of Bino, Wino and
  Higgsino.

  \begin{figure}
    \begin{center}
      \scalebox{.6}{\includegraphics{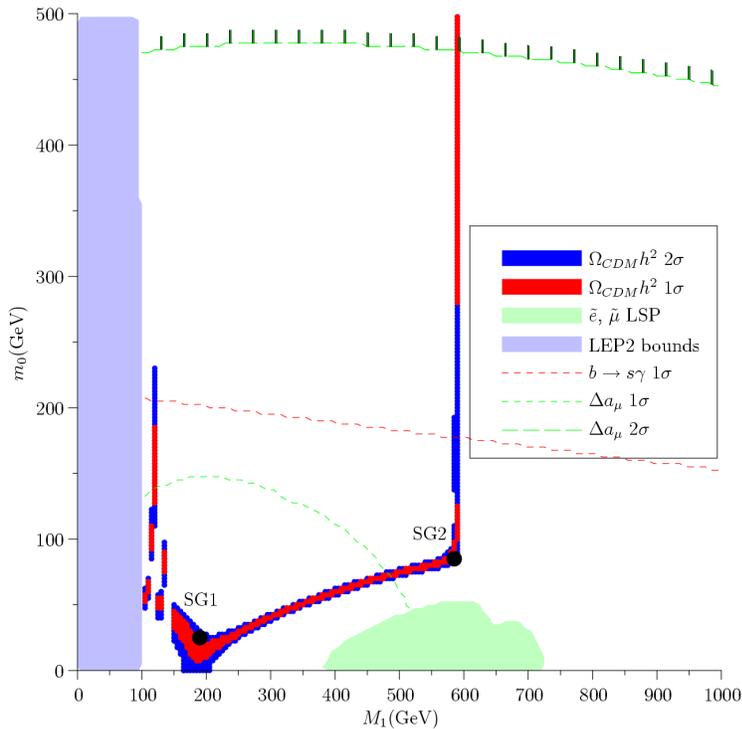}}
    \end{center}
    \caption{The $(M_1,m_0)$ plane for non-universal gaugino and
      sfermion masses with $M_2=M_3=350\ GeV$,
      $m_{0,3}=1000\ \text{GeV}$, $A_0=0$,
      $\tan\beta=10$. \label{fScalarGauge1}}
  \end{figure}

  In Fig.~\ref{fScalarGauge1} we take $M_{2,3}=350\ \text{GeV}$,
  $\tan\beta=10$ and $A_0=0$ as in \ref{fGauge1} but set
  $m_{0,3}=1\text{TeV}$. The introduction of a high 3rd family
  mass has had the same general effects as in
  Fig.~\ref{fScalar1}. The light shaded (green) excluded region is
  now due to smuon and selectron LSPs rather than staus which
  once again allows access to $m_0=0$. The lack of a light stau
  has also reduced the $\neut\neut\rightarrow l\overline{l}$
  cross-section across the entire parameter space. Thus we need
  lower values of $m_0$ and $M_1$, and thus lighter smuons and
  selectrons, to access a region in which t-channel slepton
  exchange alone can satisfy dark matter constraints.

  Apart from these details, the general features remain the same
  as in Fig.~\ref{fGauge1}. We cannot access $M_1<100\ \text{GeV}$
  due to LEP2 bounds on the lightest neutralino. At low $M_1$ we
  have thin lines corresponding to the light Higgs and Z
  resonances. The line is broken due to the resolution of the
  grid used to scan the space - a testament to extreme
  sensitivity of these regions to $M_1$. At low $M_1$ and low
  $m_0$ we have dominant contributions from t-channel slepton
  exchange. For moderate values of $M_1$ we have coannihilation
  with selectrons and smuons and when $M_1\approx 600\ \text{GeV}$,
  we once again get Bino/Wino dark matter.

  \begin{table}[ht]
    \begin{center}
      \begin{tabular}{|l|l|l|}
    \hline
    \multicolumn{1}{|c}{Parameter }&
    \multicolumn{1}{|c}{Value}&
    \multicolumn{1}{c|}{}\\
    \cline{2-3}
    &                               SG1 &         SG2     \\
    \hline
    $\Delta_{m_0}^{\Omega}$ &       $0.25$ &     $1.6$  \\
    $\Delta_{m_{0,3}}^{\Omega}$ &   $0.073$ &    $0.15$ \\
    $\Delta_{M_1}^{\Omega}$ &       $0.62$ &     $8.9$   \\
    $\Delta_{M_2}^{\Omega}$&        $0.0080$ &   $7.7$  \\
    $\Delta_{M_3}^{\Omega}$ &       $0.027$ &    $0.89$  \\
    $\Delta_{\tan\beta}^{\Omega}$ & $0.0020$ &   $0.056$ \\
    $\Delta_{A_0}^{\Omega}$ &       $0$ &        $0$ \\
    \hline
    $\Delta^{\Omega}$ &             $0.62$ &     $8.9$\\
    \hline
    $\Delta^{\text{EW}}$ &          $240$ &      $240$\\
    \hline
      \end{tabular}
      \caption{Fine tuning sensitivity at points taken from
      Fig.~\ref{fScalarGauge1} with $M_2=M_3=350\ GeV$,
      $m_{0,3}=1000\ \text{GeV}$, $A_0=0$, $\tan\beta=10$. SG1 has
      $m_0=25\ \text{GeV}$, $M_1=190\ \text{GeV}$. SG2 has
      $m_0=85\ \text{GeV}$,
      $M_1=585\ \text{GeV}$. \label{tScalarGauge1}}
    \end{center}
  \end{table}

  In our hunt for natural dark matter, we once again consider
  points in which many different annihilation channels
  contribute. We take point SG1 at the intersection of the
  selectron, smuon coannihilation region and the t-channel
  slepton band. Point SG2 is taken in the region where Bino/Wino
  dark matter also co-annihilates with selectrons and smuons.
  SG1 once again shows that regions with low $m_0$ and $M_1$
  exhibit small fine-tuning. Here the dark matter band appears at
  lower $m_0$ and lower $M_1$ in Fig.~\ref{fScalarGauge1} than in
  Fig.~\ref{fScalar1}. The result, through a combination of the
  effects discussed earlier, is an order of magnitude decrease in
  the sensitivity of this region in comparison to the most
  favourable region of the CMSSM. Point SG1 again provides a
  supernatural solution to dark matter\footnote{In this region
  we also agree with $\gmu$ in contrast to $m_0=0$ point in
  section \ref{Scalar}. This is because $M_2=350\ \text{GeV}$ which
  avoids charginos that are too light and keeps $\gmu$ at the
  correct level.}.

  \begin{figure}
    \begin{center}
      \scalebox{.6}{\includegraphics{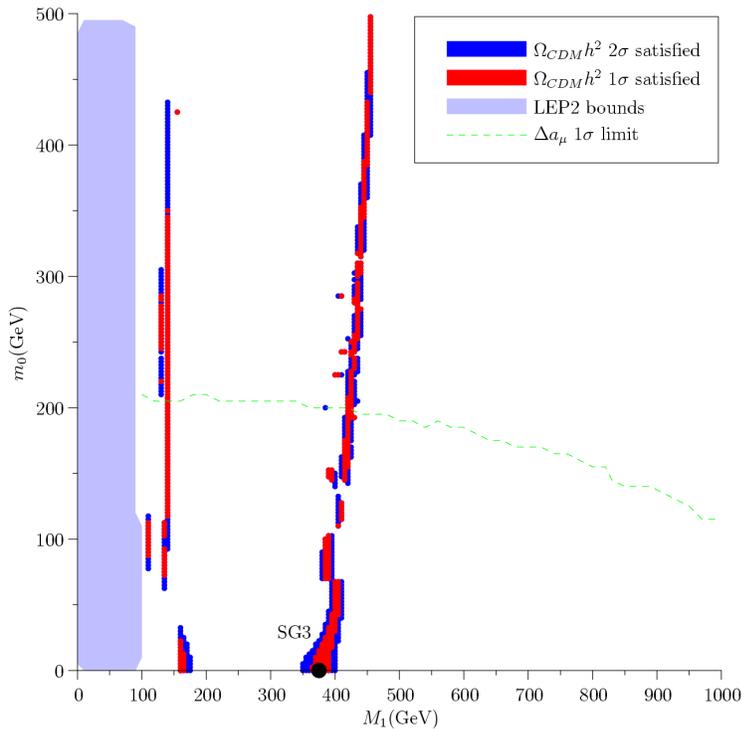}}
    \end{center}
    \caption{The $(M_1,m_0)$ plane for non-universal gaugino and
    sfermion masses with $M_2=M_3=350\ GeV$,
    $m_{0,3}=2250\ \text{GeV}$, $A_0=0$,
    $\tan\beta=10$.\label{fScalarGauge2}}
  \end{figure}

  Fig.~\ref{fScalarGauge2} contains the maximally
  tempered neutralino at point SG3. With $M_2=M_3=350\ \text{GeV}$,
  we take $m_{0,3}=2250\ \text{GeV}$ to
  achieve small enough $\mu$. Once again this raises questions of
  fine-tuning in the electroweak sector but we leave such
  concerns aside for the time being. The region of interest lies
  at $M_1\approx 400\ \text{GeV}$. This line is rather disjointed
  as {\tt SOFTSUSY} has difficulty calculating the spectrum in
  regions where $\mu$ is small.
  In this band the neutralino has significant portions of Bino,
  Wino and Higgsino. This results in all of the neutralinos and
  charginos being close in mass and so results in a large number
  of annihilation and coannihilation processes in the early
  universe. In Table~\ref{tScalarGauge2} we study point SG3 at
  $m_0=0\ \text{GeV}$, $M_1=375\ \text{GeV}$.

  \begin{table}[ht]
    \begin{center}
      \begin{tabular}{|l|l|}
    \hline
    \multicolumn{1}{|c}{Parameter }&
    \multicolumn{1}{|c|}{Value}\\
    \cline{2-2}
     & SG3\\
    \hline
    $\Delta_{m_0}^{\Omega}$ &       $0.064$ \\
    $\Delta_{m_{0,3}}^{\Omega}$ &   $3.8$ \\
    $\Delta_{M_1}^{\Omega}$ &       $2.0$ \\
    $\Delta_{M_2}^{\Omega}$ &       $0.52$ \\
    $\Delta_{M_3}^{\Omega}$ &       $3.7$ \\
    $\Delta_{\tan\beta}^{\Omega}$ & $1.0$ \\
    $\Delta_{A_0}^{\Omega}$ &       $0.015$ \\
    \hline
    $\Delta^{\Omega}$ & $3.8$\\
    \hline
    $\Delta^{\text{EW}}$ & $240$ \\
    \hline
      \end{tabular}
      \caption{Fine tuning sensitivity at the maximally tempered
      point SG3 taken from \ref{fScalarGauge2}
      with $M_2=M_3=350\ GeV$,
      $m_{0,3}=1000\ \text{GeV}$, $A_0=0$,
      $\tan\beta=10$. SG3
        has $m_0=0\ \text{GeV}$, $M_1=375\ \text{GeV}$.\label{tScalarGauge2}}
  \end{center}
  \end{table}

  The ``maximally tempered'' neutralino at point
  SG3 exhibits dramatically lower fine-tuning than either
  the Bino/Wino (G4) or Bino/Higgsino (S5) regions. By allowing the
  neutralino to be maximally mixed, we decrease the degree of
  tuning required to satisfy dark matter by an order of magnitude
  with respect to merely ``well-tempered'' cases.

  As with the case of non-universal gaugino masses alone, or just
  a high third family scalar mass, we find that there are certain
  sweet points in parameter spaces where different annihilation
  channels contribute with roughly equal strength. These regions
  are characterised by a sharp drop in the fine-tuning of soft
  parameters required to reproduce $\DM$. By allowing both
  non-universal soft gaugino masses and non-universal 3rd family
  scalar masses we find that not only are these sweet spots
  stable against further non-universalities, but also that we can
  access values of $\Delta_a^{\Omega}$ an order of magnitude smaller
  than in the CMSSM.

  \section{Conclusions}
  \label{Conc}

  We have explored regions of MSSM parameter space with
  non-universal gaugino and third family scalar masses in which
  neutralino dark matter may be implemented naturally.  In order
  to examine the relative naturalness of different regions we
  employed a dark matter fine-tuning sensitivity parameter, which
  we use in conjunction with the similarly defined sensitivity
  parameter used for EWSB. Employing these quantitative measures
  of fine-tuning we find that $\stau$-coannihilation channel in
  the CMSSM may involve as little as 25\% tuning, due to
  renormalisation group (RG) running effects.

  For non-universal third family scalar masses in which $m_{0,3}$
  is allowed to become independently large, we find that we have
  access to $m_0=0$, and hence a solution to the SUSY FCNC
  problem, with lower $\Delta_a^{\Omega}$ than in the CMSSM at
  the expense of large SUSY contributions to $(g-2)_\mu$.
  Alternatively, with non-universal third family scalar masses, we can
  also access the well tempered Higgsino/Bino region which agrees
  with $\gmu$ at $1\sigma$, but this well tempered point involves
  $3\%$ dark matter fine tuning.  We can access certain regions
  in which we have a number of different annihilation channels at
  work and these lead to lower values of dark matter fine
  tuning. However large $m_{0,3}$ implies a high degree of fine
  tuning for REWSB, as expected, so a fully natural model is not
  possible in this case.

  With non-universal gauginos the bulk region involving Bino
  annihilation via t-channel slepton exchange can be accessed,
  since low $M_1$ is possible leading to light sleptons. This
  allows supernatural dark matter (for example at points G1, G2)
  with minimal EWSB fine tuning,
  depending on how low $M_3$ is taken consistently with the LEP
  Higgs bound.  Low $m_0$ and $M_1$ also allow us to fit $\gmu$
  at $1\sigma$.  Alternatively, with non-universal gauginos,
  the well tempered Bino/Wino neutralino scenario
  is also possible, but again requires $3\%$ fine tuning
  as in the case of the well tempered Bino/Higgsino. However
  combinations of annihilation channels on the edge of such
  regions such as point G3 lead to lower fine tuning.  To achieve
  the observed value of $\DM$ through resonant annihilation such
  as at point G5 through a light Higgs requires extremely large
  fine tuning.

  With both non-universal third family masses and non-universal
  gauginos, a new feature appears: the ``maximally tempered''
  Bino/Wino/Higgsino neutralino where the LSP consists of roughly
  equal amounts of Bino, Wino and Higgsino (for example point SG3).
  Although the
  maximally tempered neutralino has quite low fine tuning, the
  large value of $m_{0,3}$ inherent in this approach always means
  that EWSB will be very fine tuned.

  In general, having non-universal third family and gaugino
  masses opens up new regions of MSSM parameter space in which
  dark matter may be implemented naturally.
  In particular allowing non-universal gauginos opens up the bulk region
  that allows Bino annihilation via t-channel slepton exchange,
  leading to ``supernatural dark matter''
  corresponding to no fine-tuning at all with respect to dark matter.
  By contrast we find that the recently proposed ``well
  tempered neutralino'' regions involve substantial fine-tuning
  of MSSM parameters in order to satisfy the dark
  matter constraints, although the fine tuning may be ameliorated
  if several annihilation channels act simultaneously. Although
  we have identified regions of ``supernatural dark matter'' in which
  there is no fine tuning to achieve successful dark matter, the
  usual MSSM fine tuning to achieve EWSB always remains.

\section*{Acknowledgements}

SFK would like to thank John Ellis and the members of the CERN Cosmo Coffee
Club for helpful comments. SFK would also like to thank Lee Roberts
for discussing the latest results relevant to the muon $g-2$ discrepancy.
JPR acknowledges a PPARC Studentship.

\end{document}